# Wavelength Controllable Forward Prediction and Inverse Design of Nanophotonic Devices Using Deep Learning

Yuchen Song[(1)], Danshi Wang[*,(1)], Han Ye[(1)], Jun Qin[(2)], and Min Zhang[(1)]

[(1)] State Key Laboratory of Information Photonics and Optical Communications, Beijing University of Posts and Telecommunications, Beijing 100876, China, [*]danshi_wang@bupt.edu.cn
[(2)] State Key Laboratory of Advanced Optical Communication Systems and Networks, Peking University, Beijing 100871, China

**Abstract** *A deep learning-based wavelength controllable forward prediction and inverse design model of nanophotonic devices is proposed. Both the target time-domain and wavelength-domain information can be utilized simultaneously, which enables multiple functions, including power splitter and wavelength demultiplexer, to be implemented efficiently and flexibly.*

## Introduction

Nanophotonic devices designed at a sub-wavelength scale attract increasing attention from various areas of computing, communications, and sensors[1]-[3] owing to their superior capability in manipulating incident electromagnetic waves. Forward response prediction of such devices is conventionally realized by the model-driven methods, which are characterized based on rigid analytical models and strongly dependent on the mathematical and physical theories, such as the finite element method (FEM) and the finite-difference time-domain method (FDTD)[4]. As the opposite process, inverse design is complex and nonintuitive due to the large design parameter space and the unclear inverse physical process. Several systematic design approaches have been proposed, including nonlinear search[5] and adjoint method[6]. However, these methods require massive forward prediction iterations, resulting in the high computational cost and noticeable performance deterioration with the increasing complexity of the device structures.

Recently, as a strong representation analysis method, deep learning (DL) has been applied into optics and photonics areas for both forward prediction and inverse design[7]-[10]. The implicit input-output relationship can be revealed by using deep neural networks (NN), so that facilitating the prediction and inverse design efficiently in a data-driven way. Silicon-based integrated nanophotonic devices with an artificially designed digital design region (DDR)[11] can implement multiple functions, including power splitter, wavelength demultiplexer, etc. In this field, however, most of the applications of DL simply performed a mapping between input and output[7], which can merely implement either forward prediction or inverse design for only one specific function. When different design targets are needed, the whole process (data collection and NN training) and NN structure have to be repeated and readjusted. Thus, compared with model-driven methods those are based on universal physical insights with various parameter controllers, the data-driven methods based on DL encounter severe challenges in flexibility and scalability. To address these problems, the prospective DL-based methods require more advanced parameter controllers to execute scalable modeling functions and embody comprehensive physical information.

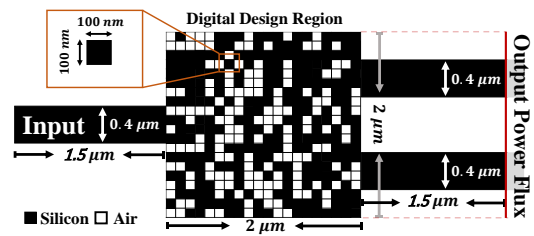

**Fig. 1:** Geometry of the investigated compact silicon-based nanophotonic device.

In this paper, a wavelength controllable data-driven approach utilizing convolutional neural networks (CNN) is proposed to implement both forward prediction and inverse design for compact DDR-based nanophotonic devices. Combined with the DDR feature extractor and wavelength controller, the proposed method can predict output power flux from arbitrary DDR in the selected wavelength range. By incorporating the wavelength controller, the characteristics from both time-domain and wavelength-domain can be taken into account in the inverse design, and thus enabling multifunctional design, which can greatly shorten the computation time and reduce the reliance on expertise. This method paves the way for hybrid parameters and data-driven modeling of nanophotonics and aims to promote the feasibility, flexibility, and scalability of DL in nanophotonics.

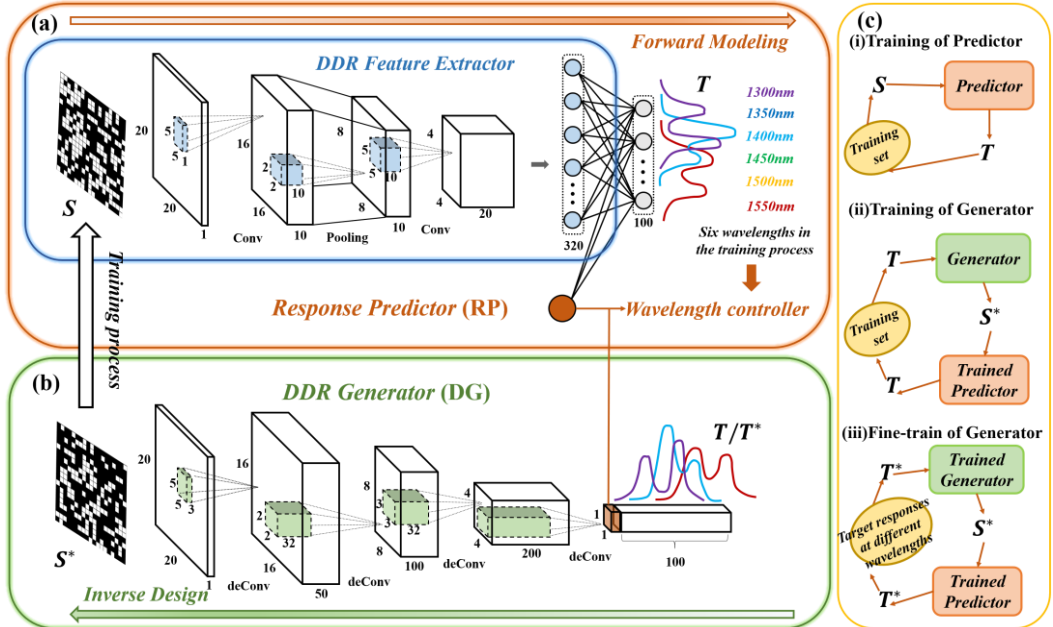

**Fig. 2:** (a) Response Predictor denoted as RP, composed of DDR Feature Extractor and wavelength controller, used to predict responses *T* at selected wavelengths from DDR *S*. (b)DDR Generator denoted as DG, used for retrieving DDR *S\** satisfying the target responses *T/T\** at specific wavelengths. (c)Three training processes of the proposed model.

**Wavelength controllable data-driven model**
In this study, we investigate a compact silicon-based nanophotonic device, as depicted in Fig. 1. A compact footprint of 2×2 µm$^2$ DDR is integrated with one input and two outputs 0.4 µm wide silicon slab waveguides. Transverse electric (TE) mode is used as input with wavelength range of 1300~1550 nm at 50 nm interval under a fixed power. The refractive indexes of silicon and air are set as 3.464 and 1. The square DDR is composed of 20×20 pixels, and each pixel with an area size of 100×100 nm$^2$ contains a binary state: "1" for non-etched silicon and "0" for etched air. The DDR represented by a 20×20 binary pixels image (denoted as *S*) and the corresponding 101 sample points of output power flux at six wavelengths (denoted as *T*) are collected as data set. A total of 8,000 data pairs are collected, 80% and 20% of which are used for training and testing, respectively. Note that *S* is generated randomly by uniform distribution.

Considering the complexity of the DDR with total $2^{20\times20}$ cases, the CNN as a powerful image information analysis technique is adopted to deal with this problem. The structure of the proposed model is composed of Response Predictor (RP) and DDR Generator (DG), as displayed in Fig. 2. The RP is first trained alone and the training of DG needs the assistance of the trained RP. Further, the DDR Feature Extractor combined with the wavelength controller make up the RP, as depicted in Fig.2 (a). The Extractor utilized CNN to compress and extract the information of DDR as a dimensionality reduction process. The wavelength information is entered through the wavelength controller and further processed along with the extracted information to predict the responses at the corresponding wavelengths. We use *S* and corresponding wavelengths as input and *T* as labels in the training data set to train the RP. Note that the responses *T* of six wavelengths ranging from 1300 to 1550nm are trained for every *S*. The RP losses of the training and testing data are around 0.04 and 0.092 at the end of training after 100 epochs.

Both the RP and FEM tools are employed to predict the corresponding responses at six wavelengths from DDR in test data. The average mean squared error (MSE) between RP and FEM from 1300 to 1550 nm is measured and shown in Fig. 3. For the purpose of visualization, the responses of the testing data at wavelengths of 1300/1325/1400/1500 nm are also displayed. It can be seen that although the responses vary significantly with different wavelengths, the responses predicted by RP coincide well with FEM results. In particular, the cases of wavelengths at 1325/1375/1425/1475/1525 nm have never appeared in the training process, but only slight mismatches occur at these untrained wavelengths, which means the wavelength controller performs good generalization ability from this training method and is similar to the role played by the wavelength parameter in the physical model-driven method. Therefore, the proposed RP is anticipated to be a supplementary and potential simulation tool in a data-driven manner.

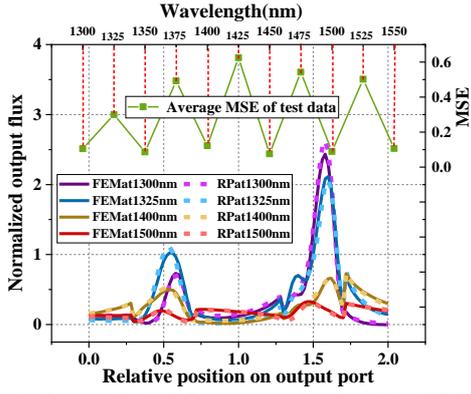

**Fig. 3:** (BottomXLeftY) Responses predicted by FEM and RP of a tested DDR data at 1300/1325/1400/1500 nm wavelength. (TopXRightY) Average testing data MSE with wavelengths from 1300 to 1550 nm at 25nm interval.

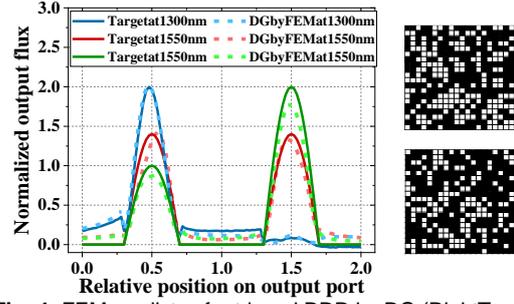

**Fig. 4:** FEM predicts of retrieved DDR by DG (RightTop: red line, RightBottom: green line) with target test and manually specific curves.

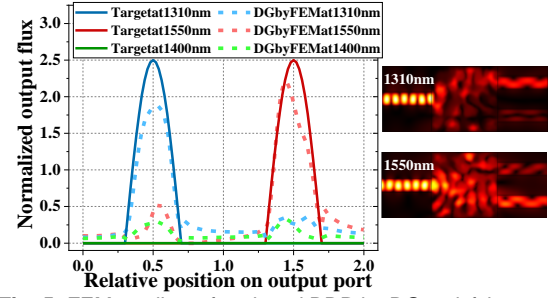

**Fig. 5:** FEM predicts of retrieved DDR by DG satisfying three target responses at different wavelengths and corresponding electromagnetic energy density at 1310/1550 nm.

Inversely, the DG is a transposed CNN to retrieve the design topology $S^*$ for the target responses $T$ with wavelengths. At training stage, the DG is cascaded in front of the trained RP. As indicated in Fig. 2(c)(ii), the input of DG are $T$ and corresponding wavelengths in the training set. The output $S^*$ of DG are fed into the trained RP, and the same $T$ act as labels at the output of RP. During this process, the parameters of RP are fixed when DG's updating. Based on this tandem structure, the contradiction that similar responses can be generated from totally different DDR can be resolved[12]. Once training is finished, DG is capable of retrieving DDR satisfying the target responses at specific wavelengths. The output $S^*$ of DG is binarized to be 0 or 1 in the end. When training DG, the test loss is around 0.14 after 100 epochs. In the whole training, the batch size is 20, the learning rate is 0.002, MSE loss function and Adam optimizer are used.

To demonstrate the feasibility of the DG, we select target responses of actual design region in the testing set as well as manually specific twin peaks curves as the input of DG. The target responses and their corresponding DDR generated by DG are shown in Fig. 4. With average MSE < 0.4, the FEM results of generated DDR are retrieved with high fidelity. In Fig 4, such twin peaks with output flux splitting ratios of 0.998 (in red line) and 0.523 (in green line) can implement 1:1 and 1:2 power splitters at desired wavelength range respectively, and other arbitrary power ratios can also be realized.

As long as the DG is trained, we can inverse design some functions considering both time and wavelength domain through it. Multiple target responses $T^*$ at different wavelengths can be used to fine-train the trained DG simultaneously as depicted in Fig.2 (c)(iii), and the DG can retrieve $S^*$ satisfying $T^*$ on the basis that the trained DG already contains inverse design information, which can not be done from scratch. Additionally, three target specific curves including single peak at corresponding port at 1310/1550 nm and zero at 1400 nm are input and the DG outputs a DDR functioned as a 1310/1550 nm wavelength demultiplexer satisfying target curves simply after 5 epochs within one minute, as shown in Fig. 5. Meanwhile, the amplitude of the target curves can be changed to obtain different efficiency results with a maximum around 90%.

**Conclusions**

In this paper, we proposed a DL-based wavelength controllable response prediction and inverse design model by incorporating the wavelength controller. Both the target time and wavelength information can be utilized simultaneously. The multiple functions of power splitters with arbitrary splitting ratio and wavelength demultiplexers at different wavelengths were implemented in a compact footprint ($2 \times 2$ μm$^2$). This study aims to improve the flexibility and controllability of DL-based method in photonics research.

**Acknowledgements**
This work was supported by National Natural Science Foundation of China (61975020, 61871415) and Fund of State Key Laboratory of IPOC (BUPT) (No. IPOC2020ZT05), P. R. China.